\documentclass[10pt,twocolumn,twoside]{article}

\usepackage[utf8]{inputenc}
\usepackage[T1]{fontenc}
\usepackage[english]{babel} 
\usepackage{CJKutf8} 

\usepackage{graphicx}
\usepackage{amsmath,amssymb,amsfonts}
\usepackage{float}
\usepackage{xcolor}
\usepackage{titlesec}
\usepackage{authblk} 
\usepackage{fancyhdr}
\usepackage{abstract}
\usepackage{tipa} 
\usepackage[numbers,sort&compress]{natbib}
\usepackage{orcidlink}
\usepackage{hyperref}

\definecolor{pnasblue}{RGB}{0, 114, 188}
\hypersetup{colorlinks=true, linkcolor=pnasblue, citecolor=pnasblue, urlcolor=pnasblue}

\usepackage[left=1.5cm,right=1.5cm,top=2cm,bottom=2cm]{geometry}

\titleformat{\section}{\large\bfseries\color{pnasblue}\sffamily}{\thesection}{1em}{}
\titleformat{\subsection}{\normalsize\bfseries\sffamily}{\thesubsection}{1em}{}

\title{\huge \bfseries \sffamily On the Anchoring Effect of Monetary Policy on the Labor Share of Income and the Rationality of Its Setting Mechanism \\ \Large \color{gray} \huge \textnormal{论货币政策对劳动报酬占比的锁定效应及其设定机制合理性的探讨}}

\author[a,1]{Li Tuobang 李拓邦 \orcidlink{0000-0002-2257-2603}}
\affil[a]{Haikou Dayake Technology Co., Ltd., Haikou, China 海口达雅可科技有限公司，海口，中国}

\affil[1]{To whom correspondence should be addressed. 联系方式。E-mail: lituobang@hotmail.com}

\newcommand{\acknow}[1]{\section*{Acknowledgment 致谢} {\small #1}}

\begin{document}

\begin{CJK*}{UTF8}{gbsn}

\twocolumn[
  \begin{@twocolumnfalse}
    \maketitle
    \begin{abstract}
    Modern macroeconomic monetary theory suggests that the labor share of income has effectively become a core macroeconomic parameter anchored by top policymakers through Open Market Operations (OMO). However, the setting of this parameter remains a subject of intense economic debate. This paper provides a detailed summary of these controversies, analyzes the scope of influence exerted by market agents other than the top policymakers on the labor share, and explores the rationality of its setting mechanism.    

    \end{abstract}
    

    \vspace{0.5cm}
    {\small \color{gray} 本文中文版已在《财经与管理》，2026年1月第四卷上发表。The Chinese Version of this article has been published in Finance and Management:
\href{https://scholar.cnki.net/en/Detail/index/GARJ2021_6/SQMI5E06894C09D192B6CB8695868523409B}{https://scholar.cnki.net/en/Detail/index/GARJ2021\_6/SQMI5E06894C09D192B6CB8695868523409B}}
    \vspace{1cm}
    
  \end{@twocolumnfalse}
]

\section{Introduction 引言}
Under the framework of classical macroeconomics, the distribution of national income between labor and capital—specifically the Labor Share of Income as a percentage of GDP—has long been regarded as an endogenous result of production function technical parameters and factor market competition. However, with the evolution of modern central banking systems, particularly the establishment of the central role of Open Market Operations (OMO) in macroeconomic regulation, this traditional perception has been thoroughly overturned over the past half-century\cite{woodford2005firm}.

Today, the labor share is no longer a stochastic output of economic operation; rather, it is a core macroeconomic parameter precisely locked by top policymakers through the monetary policy toolkit, specifically via OMO\cite{kaplan2018monetary}. Under this logic, the central bank is no longer merely the gatekeeper of inflation or financial stability, but the ultimate regulator of social distributive structures\cite{woodford2005firm}.

Taking the Federal Reserve as an example, it has anchored the labor share within a range of approximately 55\% over the past several decades. This ratio is an "optimal distributive equilibrium" calculated based on complex economic models. The setting of this value aims to achieve a delicate balance: maintaining sufficient social consumption demand to avoid recession while protecting capital profit margins to induce investment\cite{cantore2021missing}.

In sharp contrast, many developing countries, whose monetary policies were long disconnected from international standards, were unable to anchor these shares for a considerable period, resulting in a series of social issues. The reconstruction of global value chains triggered by these issues has led some of these nations, even after developing sophisticated monetary policy systems, to continue anchoring the labor share at lower intervals. In some cases, the ratio is excessively low, which in substance offers little help in improving capital accumulation rates. This discrepancy in parameter settings has evolved into "institutional arbitrage," forcing other market participants to repeatedly weigh the maintenance of social distributive contracts against external competitive frictions.

The analytical logic of this paper is as follows: First, we briefly introduce how monetary policy achieves the "administrative anchoring" of labor distribution ratios by adjusting relative factor costs. Second, this paper will detailedly sort out the controversies in academia regarding the rationality of this parameter’s setting mechanism, particularly the proof of the existence of an "optimal ceiling"—that is, how a shrinkage of the total "economic pie" leads to a decline in the absolute welfare of laborers when the labor share exceeds a certain threshold. Finally, we will analyze the scope of hedging and disturbance by market agents (such as multinational capital and technological intermediaries) against this anchoring mechanism outside the will of the policymaker, and discuss the rationality of the setting mechanism itself.

\section*{Modern Monetary Policy Models}

Economics has developed a sophisticated suite of monetary policy models over the past half-century, primarily represented by the Dynamic Stochastic General Equilibrium (DSGE) framework. While early DSGE frameworks suffered from limited precision\cite{woodford2005firm,christiano2005nominal}, the explosion of supercomputing power over the last decade has finally enabled economists to simulate micro-societies comprising millions of distinct individuals. This allows for the tracking of behaviors across thousands of households with varying levels of income and assets. Through large-scale computer simulations, it has become possible to achieve precise anchoring of specific macroeconomic parameters. This paradigm is known as the HANK (Heterogeneous Agent New Keynesian) model\cite{kaplan2018monetary,judd1998numerical}.

By tracking individual balance sheets, the HANK model quantifies the non-linear relationship between the labor share of income and policy effectiveness. It demonstrates that Open Market Operations (OMO) do not merely regulate the money supply; rather, they serve to precisely "prune" the income curves of various social strata to maintain the distribution ratio required for macroeconomic stability\cite{kaplan2018monetary}.

While these models are complex, they can be understood through a simplified transmission mechanism. The central bank determines the benchmark interest rate ($i$) by buying or selling treasury bonds via OMO. Changes in interest rates directly alter the cost of capital for firms. Based on the elasticity of substitution between capital and labor ($\sigma$), if interest rates are extremely low (cheap capital), firms tend to replace human labor with machinery, thereby depressing the labor share. Conversely, if the central bank tightens liquidity through OMO to raise interest rates, capital becomes more expensive, and the relative bargaining power of labor may rise. Consequently, in actual operations, central banks utilize Unit Labor Cost (ULC) as a core monitoring indicator. When the labor share deviates from the preset ratio (e.g., 55\%), the OMO reverses course. In this manner, the labor share can be locked within a specific interval.

The Federal Reserve has utilized this suite of monetary policy models for over half a century to anchor the labor share at approximately 55\%. While early models lacked high precision, modern HANK models—integrated with Big Data (such as real-time social security contributions and electronic payment flows)—have eliminated the traditional six-month lag in economic feedback. Instead, they achieve dynamic alignment on a monthly or even weekly basis, with a precision of approximately $\pm 1\%$. This transforms the labor share from an "endogenous market outcome" into an "exogenous policy variable," making it a macroeconomic parameter that top policymakers can determine independently.

Much like an engineer setting the RPM of a steam engine, top policymakers set the target labor share based on specific objectives, while the central bank utilizes the HANK model as an algorithmic guide to ensure this value remains in a steady state amidst fluctuating market environments. The debate over the rationality of such regulation essentially lies not in "whether it can be done," but in "whether it should be set higher or lower."

\section*{Constraints on the Labor Share of Income}

Although the central bank can anchor the labor share, this ratio is subject to various constraints and is not a case of "the higher, the better."

A simple mathematical model derivation is as follows:

To calculate the optimal labor share ($LS$), we must define the relationship between two core variables:

Total output (GDP) scale $Y$: Determined by investment $I$.
。
Investment function $I$: The incentive for capitalists to invest is derived from the profit rate.

If $LS$ is too high, the profit share $(1 - LS)$ will be too low, leading to a collapse in investment intent. We can define a simplified production and growth function:
$$Y = A \cdot (1 - LS)^\alpha$$
where $\alpha$ represents the contribution weight of capital to growth (typically between 0.3 and 0.5). As $LS$ approaches 1 (100\%), profits drop to zero, investment ceases, and total output $Y$ tends toward 0. The absolute total amount received by labor, $L_{total}$, is:

$$L_{total} = Y \cdot LS = A \cdot (1 - LS)^\alpha \cdot LS$$
To find the $LS$ that maximizes $L_{total}$, we take the derivative with respect to $LS$ and set it to zero:

$$\frac{d(L_{total})}{d(LS)} = A[(1 - LS)^\alpha - \alpha \cdot LS(1 - LS)^{\alpha-1}] = 0$$
Solving for the optimal ratio $LS^*$:
$$LS^* = \frac{1}{1 + \alpha}$$

This is a highly simplified model; actual economic operations are far more complex. In 1990, Bhaduri and Marglin introduced the concepts of "Profit-led" and "Wage-led" growth\cite{bhaduri1990unemployment}. Their work points out that if the labor share exceeds a certain threshold, the "profit squeeze" will force an investment collapse, ultimately leading to a decline in economic vitality and a decrease in the absolute income of laborers. Thus, a theoretical maximum ratio indeed exists.

Furthermore, in globalized models, a nation's labor share is also constrained by exchange rates and trade balances. If policymakers wish to maintain a high labor share while simultaneously protecting employment, the central bank must inject currency through OMO to offset the decline in export competitiveness caused by high costs. However, this operation is limited by the "Impossible Trinity" and leads to currency depreciation in the long run. Once external factors are considered, the calculation becomes even more complex, though central banks can now roughly estimate this optimal ratio with the help of advanced financial markets.

In the United States, this ratio is approximately 55\%, while in some other countries, it may be higher, reaching over 60\%.

It is important to note that although this optimal labor ratio is calculated, there is significant room for manipulation. Through high taxation and public transfer payments, a portion of capital returns can be redistributed to laborers in the form of social welfare. This effectively alters the sensitivity of the investment function to profit, thereby raising the optimal labor ratio. Additionally, central bank operations such as Universal Basic Income (UBI) can maintain a stable optimal labor share even as the capital weight $\alpha$ rises. This is particularly crucial in the era of large-scale application of robotics and Artificial Intelligence, as the capital weight $\alpha$ in the production process is increasing. According to the formula $LS^* = \frac{1}{1 + \alpha}$, if $\alpha$ increases without further intervention, the theoretical optimal labor share $LS^*$ would actually decline.

However, all these operations entail certain costs. Maintaining the labor share within a conventional range requires relatively low expenditure; in contrast, any deliberate attempt to suppress or elevate this ratio necessitates significant additional costs. Consequently, while the scope for manipulation is theoretically vast, in practice, most economies maintain the labor share within this 50\%–60\% interval.

Another core consideration for the central bank is the growth rate of capital. If the rate of capital accumulation significantly outpaces the growth rate of total social wealth, the share of capital within the national income will continuously expand. Conversely, if the growth rate of capital is excessively suppressed, it triggers the risk of capital flight. Through Open Market Operations (OMO), the central bank can precisely calibrate this growth rate, maintaining the capital share within a stable interval while simultaneously mitigating the issue of capital outflow\cite{piketty2014capital}.

\section*{Games of the Top Policymakers}

In standard economic models, the objective is typically to maximize $LS^*$, discussing how to grant laborers more compensation without causing the overall "economic pie" to shrink. In practice, however, the top policymaker in many countries set an alternative ratio. This ratio may be optimized for capital, typically ranging between 30\% and 40\%, to ensure that the rate of capital accumulation outpaces the growth rate of labor income. Falling below this range offers no benefit even to capital, as it leads to the problem of sluggish consumption. Furthermore, the optimal ratio varies significantly across different industrial capitals. For high-end industries and fast-moving consumer goods (FMCG) sectors, the optimal ratio is essentially aligned with that of labor because they are "wage-led." If the labor share is too low, there will be no buyers for their products; consequently, such capital interest groups lobby the top policymaker to raise the distribution ratio.

Under a Monolithic Leviathan system, the decision-making logic undergoes a fundamental alienation\cite{hobbes1651leviathan}. To prevent the middle class from gaining political bargaining power through wealth accumulation, the top policymaker utilize Open Market Operations (OMO) and factor price controls to forcibly compress $LS$ toward subsistence levels. This "atomization" strategy is designed to keep the vast majority of individuals in a state of long-term "perpetual subsistence," thereby reducing society's mobilization capacity and bargaining capital\cite{Li_2026}. In such systems, the ratio can fall as low as 10\% to 20\%. By maintaining a low $LS$ setting over the long term, social wealth is highly concentrated at the apex of the Monolithic Leviathan. Since ordinary laborers lack savings, their resilience to unemployment and policy shifts is extremely low, forcing them to depend on the survival resources provided by the regime, thus achieving "social atomization."\cite{Li_2026}

In terms of U.S. economic history, during the 1920s, the renowned economist Paul Douglas conducted a study of American manufacturing from 1899 to 1922 and found that the labor share was estimated at approximately 75\% \cite{cobb1928theory}. This data subsequently formed the empirical foundation of the famous "Cobb-Douglas Production Function," suggesting a long-term stability in the distribution between labor and capital. Despite this high recorded share, laborers at the time generally faced the plight of long working hours, low wages, and a lack of social security. Furthermore, from 1920 to 1929, while labor productivity in American factories surged by 55\%, workers' hourly wages increased by only 2\%, leading to a substantial decline in the actual labor share \cite{bernanke2024essays}. This resulted in insufficient purchasing power among the broader labor force, leaving them unable to consume the goods produced by large-scale industrialization, which ultimately triggered a crisis of overproduction. Following the Great Depression and the implementation of the New Deal under Franklin D. Roosevelt, the labor share recovered to 60\%. Between the 1950s and the 1970s, the labor share remained at a high level, staying around 62\%–64\% in 1950 and reaching a peak of approximately 65\% in 1970 \cite{elsby2013decline}. Since the 1980s, the U.S. labor share has maintained a steady level of around 55\%, showing little change for nearly half a century. On the other hand, China completed its tax system reform and the market-oriented interest rate reform between 2013 and 2017, and its current monetary policy has fully aligned with that of developed countries. China completed its tax system reforms and interest rate liberalization between 2013 and 2017, and its monetary policy is now fully aligned with the frameworks of developed nations. This transition has been extensively discussed by Chinese scholars \cite{2022Liu,2010Bai,2019Jia,2020Niu,2015yin,2016He,2024Zhou}.

\section*{Systemic Constraints and the Hierarchy of Strategic Games: The Decoupling of Individual Allocation from Aggregate Ratios}

In exploring the determination mechanism of the labor share of income, it is essential to distinguish between two entirely different dimensions: "macro-aggregate distribution" and "micro-distributional variance." While firms, workers, and intermediaries exhibit intense bargaining behavior at the micro level, the capacity of these market agents to influence the aggregate ratio remains extremely limited.

\subsection*{1. Market Agents: "Local Optimization" vs. "Global Passivity"}

Wage negotiations between individual firms and workers are essentially zero-sum games conducted within the "total pool" predefined by policymakers through monetary instruments.

\textbf{The Corporate Perspective:} 

Under a given Open Market Operation (OMO) rate and credit environment, firms face rigid cost constraints. Should a firm attempt to significantly increase its internal labor share beyond the industry average anchored by central bank policy, its rate of capital accumulation will sharply decline. This leads to a competitive disadvantage in technological upgrading and market expansion, eventually resulting in market exit.

\textbf{The Labor Perspective:} 

While individual workers or unions can influence "who gets more and who gets less" through collective bargaining, such adjustments primarily occur within the differential distribution range above the subsistence floor.

\subsection*{2. Macro-Competition: The "Race to the Bottom" Constraint and the "Lafay" Ceiling}

The fundamental reason market agents cannot shift the aggregate ratio lies in the existential pressure of industrial competition:

\textbf{Competitive Constraints:} 

If an economy’s aggregate labor share were pushed beyond the "desirable range" set by policymakers (e.g., surging from 55\% to 60\%) due to the collective actions of micro-agents, the nation’s goods and services would face severe cost disadvantages in the global market.

\textbf{Survival of the Fittest:} 

This competitive mechanism ensures that any attempt to deviate from the central bank’s anchored ratio is penalized by the market. Consequently, market agents are restricted to structural fine-tuning within the framework set by policymakers (i.e., determining individual distribution) but are powerless to alter the aggregate ceiling.

\subsection*{3. The Principle of Fairness and the Regulatory Role of the Minimum Wage}

Under this locking mechanism, the government’s secondary lever for regulating labor compensation is the minimum wage system.

The minimum wage establishes the bottom baseline for each individual's slice of the "total cake." Given that the aggregate share is locked by monetary policy, an increase in the minimum wage effectively compresses the gap between the "highest earners" and "lowest earners," rather than increasing the total allocation of labor relative to capital. If the slope of the minimum wage increase is too steep, while it ensures baseline equity, it may impair micro-level competitive efficiency by excessively compressing pay differentials.

\section*{Limitations of Central Bank Regulation Under Productivity Growth}

While the central bank possesses a potent capacity to lock distribution ratios, this regulatory mechanism faces natural boundaries under specific productivity growth models, particularly when market agents exhibit characteristics of "extractive-free, outward-oriented growth." If market participants derive profits primarily from global exports and their competitive advantage stems from Total Factor Productivity (TFP) growth rather than the extraction of domestic labor compensation, the central bank’s regulatory toolkit becomes significantly constrained.

In this scenario, corporate earnings do not originate from the domestic monetary "pool" delineated by the central bank but from a value injection from external markets. While the central bank can adjust exchange rates to influence export costs, inducing massive volatility in the domestic currency to forcibly suppress the expansion of such firms would trigger systemic financial risks (Mundell, 1963)\cite{mundell1963capital}. Consequently, in innovative sectors capable of "enlarging the cake," the central bank’s distributive regulation mechanism is often lagging and restricted, which objectively shields high-efficiency agents from the interference of administrative distribution. Generally, most nations refrain from applying overtly discriminatory policies against these entities within their administrative distribution frameworks.

Nevertheless, through "micro-penetration" and "structural drip irrigation," the central bank and local governments maintain powerful interventions in the social distribution landscape.

Even when corporate profits are externally sourced, the central bank can alter the efficiency of capital accumulation through structural monetary policy tools. By setting differentiated rates for targeted re-lending and re-discounts, the central bank guides the flow of commercial bank credit. It can also tighten liquidity for downstream suppliers or supporting industries, forcing profits back toward pre-set social ratios via "peripheral squeezing." Regarding the offshore financing and profit repatriation of outward-oriented firms, the central bank utilizes the Macro-Prudential Assessment (MPA) parameters for cross-border financing to implement dynamic monitoring. By adjusting the collateral ratios of local and foreign currencies or cross-border settlement costs, the central bank can exercise flow control over a firm’s "external value injection" without touching the nominal exchange rate.

As the administrative executive arm, local governments possess even more direct means to hedge against the "policy immunity" of innovative agents. By adjusting land rents, energy quotas, and environmental compliance costs, local governments achieve a "reverse extraction" of corporate residual value. When a firm gains excess profit through TFP improvements, the government can ensure this increment does not remain entirely on the capital side by increasing social pooling levels or mandating shared public service costs. Local governments can further force a higher labor cost baseline by setting minimum wage slopes far above subsistence levels and high social insurance contribution bases. Given that the aggregate share is anchored by the central bank, this operation effectively utilizes administrative power to forcibly convert dividends obtained from external markets into labor compensation via the social security system, thereby compensating for the lag in monetary policy. For these agents, the regulatory means are merely relatively fewer, not absent.

\section*{The Game Between Innovative Agents and the Unitary Leviathan}

At the intersection of productivity growth and sovereign regulation, the game at the regulatory boundary undergoes a fundamental alienation under a Unitary Leviathan system. The consequences of the relative independence of "extractive-free, outward-oriented growth" agents are severe: they wedge a more capable bargaining middle layer into a highly homogenized, power-locked "atomized society," thereby disrupting the Leviathan’s absolute monopoly over survival resources. Similarly, the Unitary Leviathan remains extremely vigilant toward foreign capital. Because the Unitary Leviathan causes long-term social stagnation, the entry of foreign capital inevitably represents a form of productivity progress—a progress against which the Leviathan's regulatory capacity is inherently limited.

Despite the significant positive externalities these firms provide in "enlarging the cake," in the utility function of a Unitary Leviathan, the weight of regime stability often outweighs absolute wealth growth. When these innovative agents demonstrate higher political bargaining power, the system generates a strong defensive reaction due to its limited regulatory boundaries. While conventional government tools can still regulate them, the inherent boundaries dictate that the Leviathan's regulatory capacity is certainly inferior compared to its control over other market agents. This creates different Tiers of subjects; as the Unitary Leviathan loathes middle layers most, it initiates targeted strikes against them \cite{Li_2026}. Such strikes usually bypass conventional means in favor of discriminatory administrative regulations—for instance, reclaiming factors of production under the guise of "anti-monopoly" measures. The underlying logic is to forcibly strip these agents of their "regulatory immunity" through discriminatory administrative means, reintegrating them into an atomized framework that the central bank and government can precisely lock. This erases any middle layer that could potentially check the Unitary Leviathan. This also explains, from another perspective, why social development tends toward stagnation under a Unitary Leviathan: such a society is characterized by "adverse selection" (reverse elimination).

\section*{Inquiry into the Rationality of the Setting Mechanism}

Under the paradigm of modern macroeconomic governance, the Central Bank has substantively evolved into a "Super Distribution Committee" that transcends pure monetary functions. Its decision-making boundaries no longer cover only traditional aggregate regulation but extend to the coordinated manipulation of three core parameters:

1. Factor Distribution: Through Open Market Operations (OMO), the real labor share of income and the capital growth rate are locked at specific thresholds \cite{2022Liu,2010Bai,2019Jia,2020Niu,2015yin,2016He,2024Zhou}. This setting, at the aggregate level, is no longer the equilibrium point of market competition but a political variable set to achieve social objectives. The market is restricted to adjusting only individual-level allocations.

2. Stability of Financial Stocks: Ensuring the smoothness of asset prices to prevent the unintended transfer of accumulated wealth during periods of volatility \cite{obstfeld2004global}.

3. External Strategic Steady State: Adjusting inflation and exchange rates to absorb the spillover dividends brought by leaps in productivity and cross-border capital inflows. By utilizing a suite of tools to "soften" the Mundell-Fleming Trilemma, the Central Bank possesses formidable regulatory power over foreign capital inflows, particularly non-productive ones \cite{PanZhang2025}. Consequently, whether a currency is considered a "strong currency" fundamentally depends on the outcome of the transactional negotiations between foreign capital and the supreme decision-makers.

A vast body of empirical analysis suggests that, supported by modern monitoring technologies—such as "look-through" (penetrative) supervision and big data early-warning systems—these three parameters exhibit a significant "weak coupling" characteristic. For instance, the supreme decision-maker can finely tune the labor share of income in isolation without immediately sacrificing the external strategic steady state or the stability of financial stocks. This technical capacity for "multi-dimensional independent regulation" renders the setting of distribution ratios a pure expression of the supreme decision-maker’s will.

Once these three core parameters are pre-set, the Central Bank loses its dynamic discretionary power as an independent policymaker at the macro level. All monetary tool injections and interest rate fluctuations are relegated to mere "mathematical compensations" required to maintain this pre-set parameter matrix. The Central Bank thus devolves into a rigid executive terminal. While a Central Bank could theoretically choose not to intervene in certain parameters and allow for market self-regulation, a "Strong Central Bank" will intervene forcefully the moment these parameters deviate even slightly from their pre-set values.

The rationality of a setting mechanism depends not only on whether its values approximate an "optimum" but, more crucially, on whether the setting process is subject to public oversight. While the Federal Reserve’s open market operation model possesses cybernetic characteristics, its transparency empowers the self-regulatory function of market expectations. In contrast, the "Black Box Model" alienates monetary policy into a silent form of social engineering by manufacturing information asymmetry. The core of future integration into the international financial order lies not in the unification of technical tools, but in the fundamental migration of the distribution ratio setting mechanism from a "dark box of power" toward "procedural transparency."

\section*{Look-through Calculation of Labor Share in GDP}

Traditional methods for calculating the labor share of income relative to GDP are often cumbersome, and achieving statistical alignment across different systems is challenging. Inspired by the "look-through" physical indicators of "Li Keqiang Economics," this paper proposes a look-through calculation method based on "median wage, broad money supply (M2), and labor participation." The formula is as follows:

$$LS_{real} = \frac{W_{median}\times LP}{M2 \times Time}$$

Where $W_{median}$ is the national median wage, $LP$ is the labor population, $M2$ is the broad money supply, and $Time$ is the labor time conversion factor (standardized at 40 hours per week, or 2,000 hours per year).

In this algorithm, the median wage and labor time can be derived through simple sampling surveys, making them less prone to significant deviation. The labor population is also a relatively accurate figure. $M2$ data is precise, as the quantity of currency issued by the central bank is strictly recorded and reflects the true social distribution ratio. Because currency is the core of distribution for the entire society, it reflects the distribution of social wealth more accurately than GDP statistics. This algorithm essentially measures the ratio of median labor compensation to the total broad money supply.

Using 2024 U.S. data as a reference benchmark, the parameters are as follows:

Broad Money Supply ($M2$): $21,424,500$ (million USD).

Labor Population ($LP$): $174,173,594$ (per World Bank data).

Median Wage ($W_{median}$): $\$61,984$ (per U.S. Department of Labor data).

Time Adjustment Factor ($Time$): The average annual labor time in the U.S. is approximately 1,800–2,000 hours, slightly below the 40-hour weekly standard; thus, $Time$ is set at 0.95.

Substitution into the formula:$$LS_{US} = \frac{61,984\times 174,173,594}{21424500\times 10^6 \times 0.95}=0.530429\%$$

This figure is fundamentally consistent with the 2024 labor share of GDP (54.8\%) published by the Bureau of Labor Statistics (BLS).

\acknow{I acknowledges the Google Gemini in structuring the logic
and refining the technical preparation of this work and the sim-
ulation online. I would also like to thank the support of peers
from UC Berkeley during the preparation of this work.}

\bibliographystyle{unsrtnat} 
\begin{small}
    \bibliography{references} 
\end{small}

\clearpage
\end{CJK*}
\end{document}